\documentclass[nofootinbib,twocolumn,preprintnumbers]{revtex4-1}
\usepackage{amsmath,amssymb,graphicx,booktabs,bm,psfrag}

\usepackage{color}

\begin{document}

\title{An Effective Field Theory for Jet Processes}

\preprint{MITP/15-065}
\preprint{August 26, 2015}

\author{Thomas Becher$^a$}
\author{Matthias Neubert$^{b,c}$}
\author{Lorena Rothen$^a$}
\author{Ding Yu Shao$^a$}

\affiliation{${}^a$Institut f\"ur Theoretische Physik {\em \&} AEC, Universit\"at Bern, Sidlerstrasse 5, CH-3012 Bern, Switzerland\\
${}^b$PRISMA Cluster of Excellence {\em \&} MITP, Johannes Gutenberg University, 55099 Mainz, Germany\\
${}^c$Department of Physics, LEPP, Cornell University, Ithaca, NY 14853, U.S.A.}

\begin{abstract}
Processes involving narrow jets receive perturbative corrections enhanced by logarithms of the jet opening angle and the ratio of the energies inside and outside the jets. Analyzing cone-jet processes in effective field theory, we find that in addition to soft and collinear fields their description requires degrees of freedom which are simultaneously soft and collinear to the jets. These collinear-soft particles can resolve individual collinear partons, leading to a complicated multi-Wilson-line structure of the associated operators at higher orders. Our effective field theory provides, for the first time, a factorization formula for a cone-jet process, which fully separates the physics at different energy scales. Its renormalization-group equations control all logarithmically enhanced higher-order terms, in particular also the non-global logarithms.
\end{abstract}
\maketitle

\section{Large logarithms in jet cross sections}
 
Jet cross sections are the most important class of observables used to study high-energy processes, because they closely mirror the underlying hard-scattering reaction. They are thus well suited to study short-distance physics and play an important role in the search for physics beyond the standard model. Such cross sections were introduced in a seminal paper by Sterman and Weinberg \cite{Sterman:1977wj}, who defined a dijet cross section in $e^+e^-$ collisions by requiring that for jet events all energy must be inside oppositely directed narrow cones of half-angle $\delta$, except for a small fraction $\beta/2$ of the center-of-mass energy $Q$. They computed the corresponding cross section as
\begin{equation}\label{SWjet}
   \frac{\sigma(\beta,\delta)}{\sigma_0} 
   = 1 + \frac{\alpha_s C_F}{4\pi} \left( - 16\ln\delta\ln\beta - 12\ln\delta + c_0 \right) ,
\end{equation}
where $\sigma_0$ is the Born-level cross section, $\alpha_s\equiv\alpha_s(\mu)$ is the strong coupling constant, and $c_0$ is a constant. Jet cross sections must include soft and collinear radiation to be infrared finite, but \eqref{SWjet} shows that there are two related problems affecting them: (i) higher-order corrections are enhanced by logarithms of $\delta$ and $\beta$, (ii) the appropriate value for the renormalization scale is unclear: should one choose $\mu=Q$, $Q\delta$, $Q\beta$ or $Q\beta\delta\,$? These difficulties are present in any perturbative computation of a multi-scale problem and the standard solution is to factorize the physics associated with the disparate scales. Once this has been achieved, one can compute each contribution at the appropriate scale and then use evolution equations to resum the large logarithms to all orders. Such resummations have been performed for many collider observables (see \cite{Luisoni:2015xha} for a recent review), but factorization and resummation for jet observables has remained an important open problem. In this letter, we will, for the first time, obtain a factorization theorem which achieves full scale separation for a jet cross section. 

A convenient method to separate the physics asocciated with different scales is to use low-energy effective field theories. Soft-Collinear Effective Theory (SCET) \cite{Bauer:2000yr,Bauer:2001yt,Beneke:2002ph} has been successfully applied to perform resummations of collider obervables \cite{Becher:2014oda}, but it has proven difficult to apply this framework to jet processes. An important stumbling block is that for such processes the usual factorization of cross sections into jet and soft functions is insufficient to achieve a complete scale separation, since the relevant soft functions suffer from large logarithms themselves. These so-called non-global logarithms (NGLs) \cite{Dasgupta:2001sh} arise when the soft radiation is not distributed evenly. This is the case for all observables involving hard phase-space cuts and also for all jet cross sections, because they are insensitive to soft radiation in the jet direction. For the jets defined in \cite{Sterman:1977wj}, NGLs arise because only a small amount of radiation is allowed outside the jets, but no restriction is imposed on the radiation inside. NGLs have been computed at two \cite{Dasgupta:2001sh,Kelley:2011ng,Hornig:2011iu,Kelley:2011aa,vonManteuffel:2013vja} and more loops \cite{Schwartz:2014wha,Khelifa-Kerfa:2015mma}, and methods for their resummation at the leading-logarithmic level exist \cite{Dasgupta:2001sh,Banfi:2002hw,Weigert:2003mm,Banfi:2010pa,Hatta:2013iba}. However, a fully factorized form of jet cross sections has not been available, despite recent progress towards this goal \cite{Caron-Huot:2015bja}. Recently, an approximate method for computing non-global observables based on an expansion in the number of sub-jets was proposed \cite{Larkoski:2015zka}, but it remains unclear whether there is a parametric suppression of higher-order terms in this expansion. Apart from this, the resummation of NGLs has not been addressed within SCET. In this letter we construct an effective field theory which factorizes such cross sections and allows for the resummation of NGLs with renormalization-group (RG) methods. This result opens the door for higher-logarithmic resummations for a wide class of observables. Given the prevalence of jet observables at the LHC there are many potential applications for our approach.

\section{Momentum regions in cone-jet processes} 

For concreteness and simplicity, we follow \cite{Sterman:1977wj} and consider the process $e^+ e^-\to 2\,{\rm jets}$. We use the thrust axis $\vec{n}$ as the jet axis and define two light-like vectors $n^\mu=(1,\vec{n})$ and $\bar{n}^\mu=(1,-\vec{n})$ along the jets. Using these vectors, we can rewrite any four-momentum in terms of its light-cone and perpendicular components. We use the thrust axis to split the event in two hemispheres and call particles with $n\cdot p<\bar{n}\cdot p $ right-moving. They are considered part of the right jet if $n\cdot p<\delta^2\,\bar{n}\cdot p$. For small values of $\delta$ this is equivalent to $\theta<2\delta$, where $\theta$ is the angle of a given particle with respect to the thrust axis. To define the jet cross section, we impose that the total energy outside the left and right jets fulfils the condition $2E_{\rm out} < Q\beta$ with $\beta\ll 1$. Except for the choice of the jet axis, our definitions are identical to the ones used in \cite{Sterman:1977wj}. Taking the thrust vector as the jet axis leads to a simpler form of the phase-space constraints and enables us to use existing two-loop results for the cone-jet soft function obtained in \cite{Kelley:2011aa,vonManteuffel:2013vja}.

As a starting point for the analysis of a multi-scale problem one needs to identify the configurations in momentum space (called ``regions'') which give non-zero contributions to the observable under consideration \cite{Smirnov:2002pj}. For a systematic effective field-theory treatment one then associates separate degrees of freedom (called ``momentum modes'') with these regions. Given the phase-space constraints in place, it is evident that momentum modes with scalings $(n\cdot p,\bar{n}\cdot p,p_\perp)$
\begin{align*}
   \mbox{collinear:} && p_c &\sim Q\,(\delta^2,1,\delta) \,, 
    & p_{\bar c} &\sim Q\,(1,\delta^2,\delta) \nonumber\\
   \mbox{soft:} && p_s &\sim Q\,(\beta,\beta,\beta)
\end{align*}
are needed to describe the dynamics at small values of $\beta$ and $\delta$. We restrict ourselves to the case $\beta\ll\delta$ to avoid that collinear particles recoil against soft ones. To achieve this one can choose $\beta\sim\delta^2$, in which case the soft modes become the standard ultra-soft modes of SCET$_I$.

In principle, these soft and collinear modes are sufficient to describe the jet cross section, as was shown for Sterman-Weinberg jets at one-loop order in \cite{Cheung:2009sg}. However, the soft function contains large logarithms of $\delta$ and $\beta$, which cannot be avoided by any choice of $\mu$. Such a treatment can therefore not achieve the goal of resumming the large logarithms in the cross section. We find that this problem arises because the soft function receives contributions from two different, hierarchical scales $Q\beta$ and $Q\delta\beta$, while a proper effective field theory should separate the physics at these two scales. A related problem is that the jet functions need  non-trivial (``zero-bin'') subtractions to avoid double counting. We argue that this soft-collinear overlap region contains nontrivial physics, which should be factorized. 

To avoid the presence of multiple scales in individual functions it is necessary to perform a strict multipole expansion in the effective theory: all power-suppressed contributions must be expanded away, {\em including those in the phase-space constraints.} After this expansion, the collinear particles are {\em always inside\/} the jet, since they have parametrically large energy $\bar{n}\cdot p_c\sim Q\gg Q\beta$. Likewise, the soft particles are {\em always outside\/} the cone, since they generically have large angle $n\cdot p_s/\bar{n}\cdot p_s\sim 1\gg\delta^2$. It is physically clear that there also must be contributions which involve both small energy and small angle. Indeed, we find that particles which are simultaneously collinear and soft (in short ``coft'', subscript ``$t$'')
\begin{align*}
   \mbox{coft:} && p_t &\sim Q\beta\,(\delta^2,1,\delta) \,, 
   & p_{\bar t} &\sim Q\beta\,(1,\delta^2,\delta)
\end{align*}
do give leading contributions to the cross section. Soft-collinear modes have arisen in other contexts in SCET, see e.g.\ \cite{Becher:2003qh,Bauer:2011uc,Larkoski:2015zka}. However, the fact that all components of the coft mode are smaller than the corresponding collinear ones is new and gives rise to fundamentally different interactions. Coft modes can be emitted inside or outside a jet and their natural scale is $\sqrt{p_t^2}\sim Q\delta\beta$, which is much lower than both the collinear scale $Q\delta$ and the soft scale $Q\beta$. The physical relevance of this scale is an important new result of our analysis.

After expanding the phase-space constraint in the different momentum regions (see supplemental material), it is a simple exercise to verify that one reproduces the one-loop cone-jet rate by expanding the dijet cross section in the above momentum regions, performing the phase-space integrals in each region, and adding up the resulting contributions. Integrating over the gluon phase space we find for the one-loop corrections from the different sectors:
\begin{align}\label{eq:oneloop}
   \Delta\sigma_h &= \frac{\alpha_s C_F}{4\pi}\,\sigma_0
    \left( \frac{\mu}{Q} \right)^{2\epsilon} 
    \left( -\frac{4}{\epsilon^2} - \frac{6}{\epsilon} - 16 + \frac{7\pi^2}{3}  \right) \nonumber\\
   \Delta\sigma_{c+\bar{c}} &=\frac{\alpha_s C_F}{4\pi}\,\sigma_0
    \left( \frac{\mu}{Q\delta} \right)^{2\epsilon} 
    \left( \frac{4}{\epsilon^2} + \frac{6}{\epsilon} + 16 - \frac{5\pi^2}{3} + c_0 \right) 
    \nonumber\\
   \Delta\sigma_s &=\frac{\alpha_s C_F}{4\pi}\,\sigma_0
    \left( \frac{\mu}{Q\beta} \right)^{2\epsilon}
    \left( \frac{4}{\epsilon^2} - \pi^2   \right) \nonumber\\
   \Delta\sigma_{t+\bar{t}} &=\frac{\alpha_s C_F}{4\pi}\,\sigma_0
    \left( \frac{\mu}{Q\delta\beta} \right)^{2\epsilon} 
    \left( -\frac{4}{\epsilon^2} + \frac{\pi^2}{3}  \right) 
\end{align}
where  $d=4-2\epsilon$. In the sum of the above contributions the divergences cancel and we reproduce the full QCD result given in \eqref{SWjet}, with $c_0=-2+12\ln 2$ for thrust-axis cone jets (the original Sterman-Weinberg jets have $c_0=10-4\pi^2/3$). Our collinear result is the same as the zero-bin subtracted collinear contribution obtained in \cite{Cheung:2009sg}, and the sum of our soft and coft contributions is equal to the soft result in this reference. Importantly, however, our result systematically disentangles the different scales, and our computations are much simpler because the multi-pole expansion simplifies the phase-space constraints and makes subtractions of overlap contributions unnecessary.

Given the above result, one expects that the cross section can be factorized into a product of a hard function, jet functions, and a convolution of soft and coft functions. On a basic level this is true, but the interplay between coft and collinear partons leads to a highly non-trivial structure of the corresponding factorization formula.

\section{Factorization of the cross section}
 
At first sight, the factorization of collinear and coft contributions seems to be a trivial matter. Since every single momentum component of a coft field is smaller than the corresponding component of a collinear field, we can treat coft modes as submodes of collinear fields. In other words, we can construct the relevant effective Lagrangian and operators starting from the purely collinear case and splitting the fields as $\phi_c\to\phi_c+\phi_t$. However, because {\em all\/} components of the coft fields are power suppressed compared to their collinear counterparts, there are no coft-collinear interactions in the Lagrangian: ${\cal L}_{c+t} = {\cal L}_{c}+{\cal L}_{t}$. The only place where the coft field appears is in the collinear Wilson line, which factorizes into a product of a collinear and a coft Wilson line, $W_c\to W_c\,U(\bar{n})$. The quantity $U(\bar{n})$ is defined exactly as $W_c$ but with the coft gluon field instead of the collinear one. Since we will encounter coft Wilson lines along different directions, we have explicitly included the vector $\bar{n}$ as an argument.

An important second source of coft-collinear interactions arises from on-shell collinear particles in the final state. For a coft particle with momentum $k$, emitted from a generic collinear particle with momentum $p_1$, we would approximate the propagator denominator before the emission as $(p_1+k)^2\to p_1^2$. As discussed above, coft emissions from generic collinear particles are a power-suppressed effect. However, if the virtuality of the collinear quark is zero, the leading contribution is obtained from $(p_1+k)^2\to 2p_1\cdot k$. Computing the relevant amplitude, one finds that it is equal to the gluon emission from a coft Wilson line $U(n_1)$ along the direction $n_1^\mu=2p_1^\mu/\bar{n}\cdot p_1$ of the collinear final-state particle. Repeating the computation with two  gluons, we find that the corresponding matrix element is indeed the two-gluon matrix element of the same operator.

\begin{figure}[t!]
\centering
\begin{tabular}{cc}
\includegraphics[width=.22\textwidth]{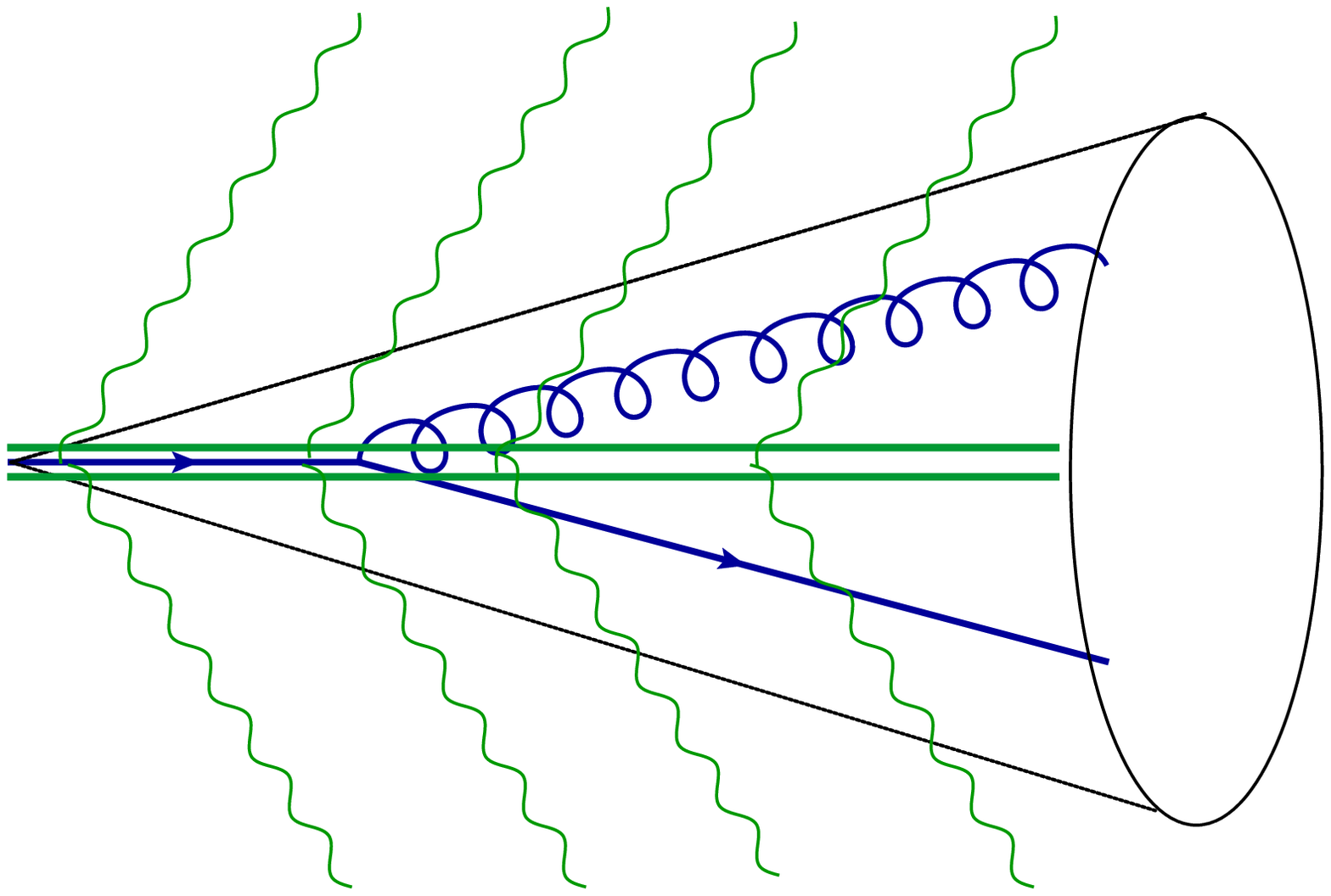} & \includegraphics[width=.25\textwidth]{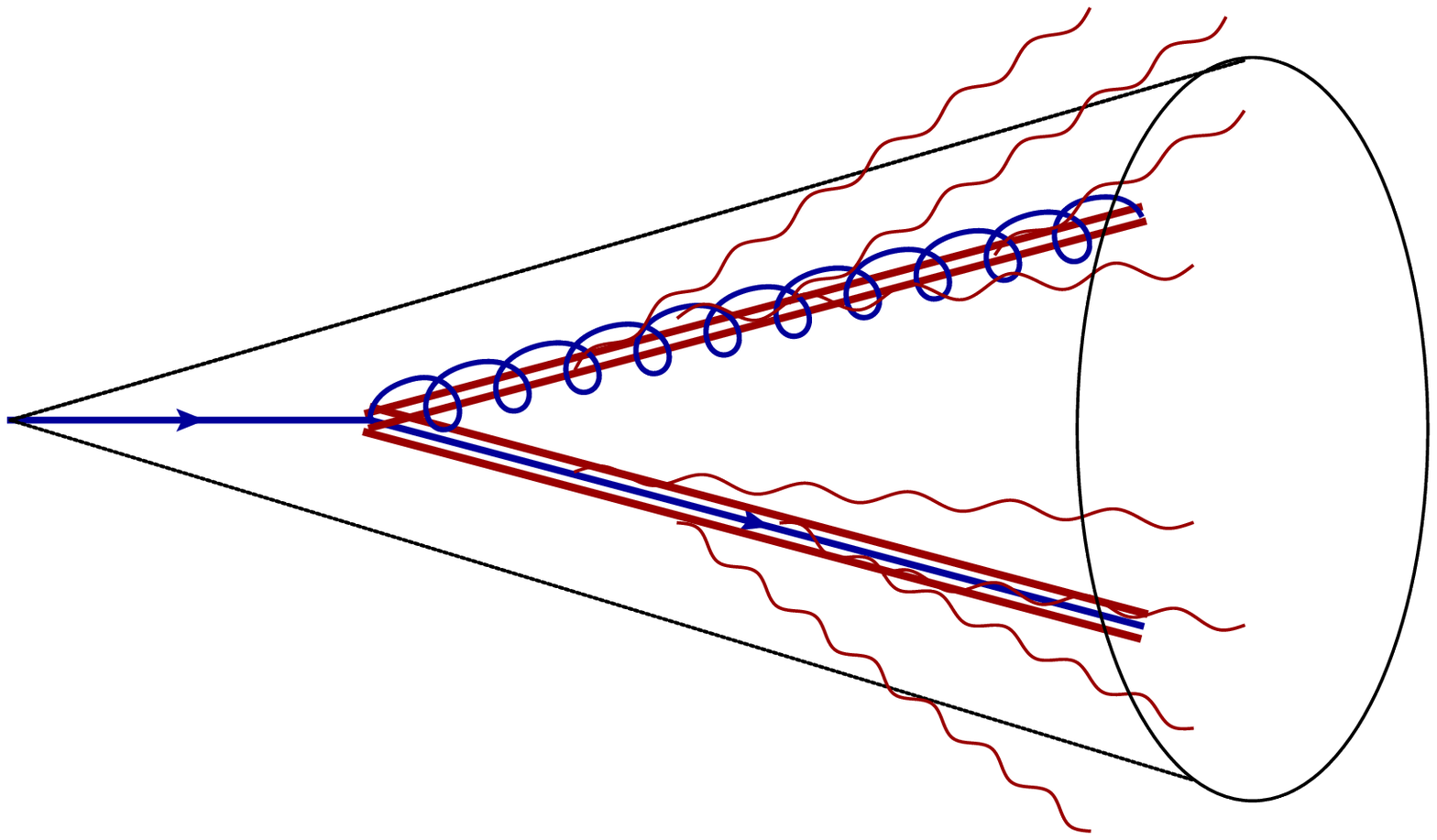} 
\end{tabular}
\caption{\label{fig:softvscoft} 
Soft factorization (left) versus coft factorization (right). Collinear particles are shown in blue, soft emissions in green and the small-angle soft radiation described by the coft mode in red. The double lines indicate the direction of the associated Wilson lines.}
\end{figure}

For a single collinear quark in the final state $n_1^\mu=n^\mu$, and the coft function is given by two Wilson lines, as would be the case for soft emissions. To see the physics difference between soft and coft modes one needs to consider the case with several collinear particles inside the jet. Doing so, one finds that every collinear final-state particle gets dressed by a coft Wilson line. In color-space notation \cite{Catani:1996vz}, the coft emissions in the presence of a final state with $m$ collinear particles can be obtained by taking the matrix element of the operator 
\begin{equation}\label{eq:Wilson}
   \bm{U}_0(\bar{n})\,\bm{U}_1(n_1)\dots {\bm U}_m(n_m)|\mathcal{M}_m(p_0;\{\underline{p}\})\rangle \,,
\end{equation}
where $|\mathcal{M}_m\rangle$ is the amplitude for the collinear quark field with momentum $p_0 \approx Q\,n/2$ to split into particles with momenta $\{\underline{p}\}=\{p_1,\dots, p_m\}$, and $\bm{U}_i(n_i)$ is a Wilson line along the direction $n_i=p_i/E_i$ in the color representation relevant for the given particle. The fact that soft emissions build up Wilson lines is of course very familiar. What is special in the present case is that the coft particles are emitted in a narrow cone and can therefore resolve the individual collinear partons. As a consequence, we end up with individual Wilson lines for each of the collinear final-state partons, instead of just one overall Wilson line describing all soft emissions, see Figure~\ref{fig:softvscoft}.  

To write down a factorized form of the cross section based on the result \eqref{eq:Wilson}, we first perform a Laplace transformation with respect to $\beta$, i.e.\
\begin{equation}
   \widetilde\sigma(\tau) = \int_0^{\infty}\!\!d\beta\,e^{-\beta/(\tau e^{\gamma_E})}\,
    \frac{d\sigma}{d\beta} \,.
 \end{equation}
This is convenient, since the energy emitted outside the cones is shared among the soft and coft degrees of freedom. The Laplace transformation factorizes the corresponding phase-space constraint. Since the cone constraint acts on the individual partons, it trivially factorizes. In Laplace space we then obtain the factorization formula
\begin{equation}\label{eq:factthm}
   \widetilde\sigma(\tau) = \sigma_0\,H(Q)\,\widetilde{S}(Q\tau)\! \left[ \sum_{m=1}^{\infty} 
    \left\langle \bm{\mathcal{J}}_{\!\!m}(Q\delta) \!\otimes 
    \widetilde{\bm{\mathcal{U}}}_m(Q\delta\tau) \right\rangle \right]^2
\end{equation}
for the jet cross section, where the angle brackets denote the color trace $\langle M\rangle=\frac{1}{N_c}\,{\rm tr}(M)$. The jet functions $\bm{\mathcal{J}}_{\!\!m}(Q\delta)$ and the coft functions $\widetilde{\bm{\mathcal{U}}}_m(Q\delta\tau)$ are obtained from squaring the amplitude \eqref{eq:Wilson}. Both depend on the directions $n_i$ of the collinear partons. The symbol $\otimes$ indicates that the product of the jet and coft functions needs to be integrated over the directions of the vectors $n_i$, and the square in \eqref{eq:factthm} takes into account the identical contributions of the left and right cone jets. Relation (\ref{eq:factthm}) is our most important result, as it is the first complete factorization formula for a cone-jet cross section. Its structure implies that simpler formulas with only one soft Wilson line per jet (see e.g.\ \cite{Cheung:2009sg,Ellis:2010rwa}) do not hold beyond one-loop order. $H(Q)$ is the familiar hard function for two-jet processes. The soft function $S(Q\beta)$ is the squared matrix element of two Wilson lines along the jet directions, with a constraint on the energy but no angle constraint, as explained earlier. The same soft function arises in threshold resummation for Drell-Yan production, up to the fact that the Wilson lines are now outgoing. This does not change the perturbative result, which at two loops was obtained in~\cite{Belitsky:1998tc,Becher:2007ty}.

The coft function with $m$ Wilson lines is given by
\begin{multline}\label{eq:Un}
   \bm{\mathcal{U}}_m(Q\delta \beta) = \int\limits_{X_t}\hspace{-0.52cm} \sum\,
   \langle 0|\bm{U}_0^\dagger(\bar{n})\,\bm{U}_1^\dagger(n_1)\dots {\bm U}_m^\dagger(n_m) 
    |X_t\rangle \\[-0.5mm]
   \times \langle X_t| \bm{U}_0(\bar{n})\dots {\bm U}_m(n_m) |0\rangle\, 
    \delta(Q\beta-\bar{n}\cdot p_{X^{\rm out}_{t}} ) \,,
\end{multline}
and the jet function containing $m$ partons is defined as
\begin{multline}
   \frac{n\!\!\!/}{2}\,\bm{\mathcal{J}}_{\!\!m}(Q\delta) 
    = 2(2\pi)^{d-1} \prod_{i=1}^m \int\!\frac{dE_i\,E_i^{d-3}}{2(2\pi)^{d-1}}\,
     \theta\Big(\delta^2-\frac{n\cdot p_i}{\bar{n}\cdot p_i}\Big)\quad \\
   \times \delta(Q-\bar{n}\cdot p_{X_c})\,\delta^{d-2}(p_{X_c}^\perp)\!\sum_{\rm spins} 
    |\mathcal{M}_m\rangle \langle\mathcal{M}_m| \,. ~
\end{multline}
The jet functions suffer from singularities when particles become soft and collinear and are therefore distribution-valued in the angles formed by the vectors $n,\bar n$ and $n_i$.

We have derived all ingredients needed to evaluate the factorization formula \eqref{eq:factthm} at two-loop order. This provides a highly nontrivial consistency check of our framework, since the individual contributions diverge as strongly as $1/\epsilon^4$ for $\epsilon\to 0$ and scale differently, cf.\ \eqref{eq:oneloop}. These divergences cancel in the final result, which can then be compared to numerical results obtained by running a fixed-order event generator such as {\sc Event2} \cite{Catani:1996vz} at small values of $\beta$ and $\delta$, finding complete agreement \cite{longer}. For completeness, we supply the explicit two-loop expressions for all relevant functions as supplemental material.

\section{Renormalization and resummation} 

The factorization formula (\ref{eq:factthm}) achieves a complete scale separation. Once the individual functions have been renormalized one can evaluate each ingredient at its natural scale and then evolve them to a common factorization scale by solving RG evolution equations, thereby resumming all large logarithms of $\beta$ and $\delta$ in the cross section. The renormalization of the hard and soft functions is multiplicative and the relevant anomalous dimensions are known to three-loop accuracy. The renormalization of the jet functions, on the other hand, cannot be multiplicative, since $\bm{\mathcal{J}}_{\!\!m}$ starts at $\mathcal{O}(\alpha_s^m)$ and contains divergences. These real-emission divergences arise from degenerate parton configurations and cancel against virtual corrections present in lower-point amplitudes. This implies that the $Z$ factor relating the bare and renormalized jet functions, defined as (summed over $k\le m$)
\begin{equation}
   \bm{\mathcal{J}}_{\!\!m}(Q\delta,\epsilon) 
   = \bm{\mathcal{J}}_{\!\!k}(Q\delta,\mu)\,\bm{Z}^{J}_{km}(Q\delta,\epsilon,\mu) \,,
\end{equation}
is an upper triangular matrix with a hierarchical structure, i.e.\ the off-diagonal elements are suppressed by powers $\bm{Z}^J_{km}\sim\alpha_s^{m-k}$. The matrix elements depend on the directions of the partons in $\bm{\mathcal{J}}_{\!\!m}$ and act on their color indices. The jet-function renormalization factor contains logarithmic dependence on the jet scale $Q\delta$, as is typical for Sudakov problems. 

Having renormalized all other elements of the factorization formula, one must now find that the matrix
$\bm{Z}^U\equiv Z_H^{1/2} Z_S^{1/2} \bm{Z}^J$ renormalizes the coft functions, i.e.\ 
\begin{equation}\label{eq:renCond}
   \widetilde{\bm{\mathcal{U}}}_k(Q\delta\tau,\mu) 
   = \bm{Z}^U_{km}(Q\delta\tau,\epsilon,\mu)\,
    \hat\otimes\,\,\widetilde{\bm{\mathcal{U}}}_m(Q\delta\tau,\epsilon) \,.
\end{equation}
This relation has several interesting features. First, it implies that the Sudakov logarithms in the various $Z$ factors must conspire to produce a dependence on only the coft scale, once $\bm{Z}^U$ is applied to the coft functions. A second, interesting feature of the matrix structure is that higher-multiplicity coft functions enter the renormalization of the lower ones. For example, the two-loop renormalization of the coft function with two Wilson lines has the form
\begin{equation}\label{eq:Uren}
   \widetilde{\bm{\mathcal{U}}}_1(\mu) 
   = \bm{Z}^U_{11}\,\widetilde{\bm{\mathcal{U}}}_1(\epsilon) 
    + \bm{Z}^U_{12}\,\hat{\otimes}\,\,\widetilde{\bm{\mathcal{U}}}_2(\epsilon) 
    + \bm{Z}^U_{13}\,\hat{\otimes}\,\bm{1} + \mathcal{O}(\alpha_s^3) \,,
\end{equation}
where we have used the fact that $\widetilde{\bm{\mathcal{U}}}_3=\bm{1}+\mathcal{O}(\alpha_s)$. The off-diagonal contributions depend on the directions of the additional partons, and the symbol $\hat{\otimes}$ indicates that one has to integrate over these since the renormalized function $\widetilde{\bm{\mathcal{U}}}_1(\mu)$ multiplies the jet function $\bm{\mathcal{J}}_{\!\!1}(\mu)$, which does not depend on these additional degrees of freedom. The renormalization condition \eqref{eq:renCond} is at first sight surprising, because Wilson-line matrix elements can usually be renormalized multiplicatively. However, we have checked explicitly that this condition renormalizes the function $\widetilde{\bm{\mathcal{U}}}_1(\mu)$ correctly to two-loop order. 

The resummation of large logarithms in the factorized cross section is accomplished by evolving the various functions to a common scale. If for convenience one chooses the jet scale as a reference, $\mu_c\sim Q\delta$, one needs the well-known solutions for the hard and soft functions along with the solution $\widetilde{\bm{\mathcal{U}}}_k(\mu_c)=\bm{U}_{km}(\mu_c,\mu_t)\,\hat\otimes\,\,\widetilde{\bm{\mathcal{U}}}_m(\mu_t)$ with
\begin{equation}\label{Uevol}
   \bm{U}(\mu_c,\mu_t) = \mbox{\bf P} \exp\left[ \int_{\alpha_s(\mu_t)}^{\alpha_s(\mu_c)}\!d\alpha\,
    \frac{\bm{\Gamma}^U(\alpha)}{\beta(\alpha)} \right]
\end{equation}
for the coft functions, where $\mu_t\sim Q\delta\tau$. The anomalous dimension $\bm{\Gamma}^U$ is defined in the usual way in terms of the $1/\epsilon$ pole terms in $\bm{Z}^U$. For the resummation at leading order in RG-improved perturbation theory one needs tree-level matching conditions and one-loop anomalous dimensions (two-loop accuracy for the cusp part, which can be factored out). At tree level only the jet-function $\bm{\mathcal{J}}_{\!\!1}$ is nonzero, while all Wilson-line matrix elements are trivially given by $\widetilde{\bm{\mathcal{U}}}_m=\bm{1}$. To this accuracy, the anomalous-dimension matrix only has entries in the diagonal and above the diagonal, $\bm{V}_m=\bm{\Gamma}^U_{mm}$ and $\bm{R}_m=\bm{\Gamma}^U_{m(m+1)}$. Our formalism applies for an arbitrary number of colors and allows one to systematically include higher logarithms. A resummation at next-to-leading order would require using one-loop matching conditions for the jet and coft functions along with two-loop anomalous dimensions (three-loop for the cusp part), in which case the jet function $\bm{\mathcal{J}}_{\!\!2}$ and the off-diagonal elements $\bm{\Gamma}_{m(m+2)}^U$ contribute as well. The upper triangular structure of the infinite-dimensional anomalous-dimension matrix $\bm{\Gamma}^U$ is responsible for the intrinsic complexity of NGLs, which prevents one from obtaining a closed analytic expression for the evolution function (\ref{Uevol}) even at leading order. However, when the path-ordered exponential is expanded in a power series, one finds an iterative structure of real emissions $\bm{R}_m$ and virtual corrections $\bm{V}_m$ resembling that of a parton shower. It is thus likely that an approximate numerical solution can be obtained using Monte Carlo techniques. In \cite{longer} we will analyze the case of wide-angle jets with $\delta\sim 1$ in detail. Since there are no collinear singularities in this case, one can use the soft limit to write down explicit expressions for $\bm{R}_m$ and $\bm{V}_m$. The corresponding one-loop anomalous-dimension matrix has a close connection to the anomalous dimension governing the functional RG equation proposed in \cite{Caron-Huot:2015bja}. We have verified that in the large-$N_c$ limit the first three terms in the power expansion of (\ref{Uevol}) agree with the corresponding expansion of the BMS equation \cite{Banfi:2002hw} derived in~\cite{Schwartz:2014wha}. 

In summary, our analysis provides, for the first time, an all-order factorization formula (\ref{eq:factthm}) for a process with NGLs and demonstrates that also this class of logarithms can be captured using RG methods. A key element of the factorization formula is small-angle soft radiation and its associated physical scale $Q\delta\beta$, which is parametrically smaller than the soft and collinear scales. A short-distance treatment of jet processes is only possible if $Q\delta\beta\gg\Lambda_{\rm QCD}$. The multi-Wilson-line structure of this radiation seems to be a generic feature of such observables. In the future, it will be interesting and important to apply our approach to other non-global observables, in particular to jet processes at hadron colliders.

\acknowledgements
T.B.~and L.R.\ are supported by the Swiss National Science Foundation under grant 153294. M.N.~is supported by the ERC Advanced Grant EFT4LHC, the PRISMA Cluster of Excellence EXC 1098 and grant 05H12UME of the German Federal Ministry for Education and Research. The authors thank the high-energy theory groups at Harvard (US Department of Energy grant DE-SC0013607) and Heidelberg, the MITP Mainz and the INT Seattle for hospitality and support.

\onecolumngrid
\newpage
\appendix

\section*{Supplemental material}

Below, we give the explicit form of the expanded phase-space constraints used to obtain the result in Eq.~(2) of the main text. Furthermore, we present the explicit expressions for the ingredients of the factorization formula (5) to $\mathcal{O}(\alpha_s^2)$ and give the result for the cone-jet cross section at next-to-next-to-leading order (NNLO).

\subsection*{A: Multipole expansion of the phase-space constraint}
\renewcommand{\theequation}{A\arabic{equation}}
\setcounter{equation}{0}

As stressed in the main text, to achieve scale separation it is crucial to systematically expand away power-suppressed terms in effective-theory computations. This multipole expansion must also be applied to the phase-space measure. In the following, we expand the phase-space measure relevant for the cone-jet cross section. Since collinear particles have parametrically large energies,  $\bar{n}\cdot p_c\sim Q\gg Q\beta$, they can never be outside the jet. The soft particles, on the other hand, generically have large angle $n\cdot p_s/\bar{n}\cdot p_s\sim 1\gg\delta^2$, so that the out-of-jet constraint is always fulfilled after the expansion. Since the method-of-regions expansion is performed on the level of the integrand, this implies that the soft particles are regarded as outside the jet, irrespective of the angle of the actual emission. Given these scalings, we can now write down the explicit form of the expanded phase-space constraint for the jet cross section, including the momentum conservation $\delta$-function. It reads 
\begin{multline}\label{eq:momcons}
   \delta(Q-\bar{n}\cdot p_{X_c})\,\delta^{d-2}(p_{X_c}^\perp)\,\delta(Q-n\cdot p_{X_{\bar c}})\,
    \delta^{d-2}(p_{X_{\bar{c}}}^\perp)\, 
    \theta(Q\beta-2E_{X_s}-\bar{n}\cdot p_{X^{\rm out}_{t}}-n\cdot p_{X^{\rm out}_{\bar{t}}}) \\
   \times \mbox{$\prod_i$}\,\theta(\delta^2 \bar{n}\cdot p_{c}^i-n\cdot p_{c}^i)\,\,  
    \mbox{$\prod_j$}\,\theta(\delta^2 n\cdot p_{\bar{c}}^j-\bar{n}\cdot p_{\bar{c}}^j) \,,
\end{multline}
where $p_{X_{\bar c}}$ is the total momentum of the collinear particles, etc. The separate constraints on the transverse momentum in each hemisphere ensure that $\vec{n}$ is indeed the thrust axis, see e.g.\ \cite{Becher:2015gsa}. The soft and coft momenta are not constrained by momentum conservation, since they are parametrically smaller than the collinear momenta. The angle constraints in the last line enforce that all collinear particles are inside the jets. As stated above, there are no angle constraints on the soft particles. The right-moving coft particles can be inside or outside the right jet, and the energy constraint in \eqref{eq:momcons} acts on them if they are outside the right jet. These right-moving coft particles do not see the left jet, because after the multipole expansion they are always outside this jet. 

\subsection*{B: Two-loop results for the bare functions}
\renewcommand{\theequation}{B\arabic{equation}}
\setcounter{equation}{0}

We first present the bare ingredients for the factorized NNLO cross section. We write our results in terms of the bare coupling constant $\alpha_0=Z_\alpha\alpha_s$, with
\begin{equation}
   Z_\alpha = 1- \frac{\alpha_s}{4\pi} \frac{\beta_0}{\epsilon} + \dots \,, 
   \quad \text{and}\quad \beta_0 = \frac{11}{3}\,C_A - \frac{4}{3}\,T_F n_f \,.
\end{equation}
Writing the hard function as a function of the logarithm $L=\ln\frac{Q}{\mu}$, we find \cite{Becher:2006mr}
\begin{align}
   H_{\rm bare}(L,\epsilon) &= 1 + \frac{\alpha_0 C_F}{4\pi}\,e^{-2\epsilon L} \left[ 
    - \frac{4}{\epsilon ^2} - \frac{6}{\epsilon} - 16 + \frac{7\pi^2}{3}  
    + \epsilon\left( - 32 + \frac{7\pi^2}{2} + \frac{28\zeta_3}{3}   \right) 
    + \epsilon^2\left(  - 64 + \frac{28\pi^2}{3}  + 14\zeta_3 - \frac{73\pi^4}{360}   \right) \right]
    \nonumber\\
   &\quad\mbox{}+ \left( \frac{\alpha_0}{4\pi} \right)^2 e^{-4\epsilon L} 
    \left(  C_F^2 h_F + C_F C_A h_A + C_F T_F n_f h_f \right) ,
\end{align}
with 
\begin{align}
   h_F &= \frac{8}{\epsilon^4} + \frac{24}{\epsilon^3}
    + \frac{1}{\epsilon^2} \left( 82 - \frac{28\pi^2}{3} \right)
    + \frac{1}{\epsilon} \left(  \frac{445}{2}  - 26\pi^2 - \frac{184\zeta_3}{3}  \right)
    + \frac{2303}{4} - 86\pi^2 - 172\zeta_3+ \frac{137\pi^4}{45}     \,, \nonumber\\
   h_A &= - \frac{11}{3\epsilon^3} + \frac{1}{\epsilon^2} 
    \left( - \frac{166}{9} + \frac{\pi^2}{3}  \right)
    + \frac{1}{\epsilon} \left( - \frac{4129}{54} +  \frac{121\pi^2}{18} + 26\zeta_3  \right)
    - \frac{89173}{324}  + \frac{877\pi^2}{27}   + \frac{934\zeta_3}{9}- \frac{8\pi^4}{45}  \,,
    \nonumber\\
   h_f &= \frac{4}{3\epsilon^3} + \frac{56}{9\epsilon^2} 
    + \frac{1}{\epsilon} \left( \frac{706}{27} - \frac{22\pi^2}{9} \right)  + \frac{7541}{81}
     - \frac{308\pi^2}{27} - \frac{104\zeta_3}{9} \,.
\end{align}
For the Laplace transformed soft function the relevant logarithm is $L=\ln\frac{Q\tau}{\mu}$. We have \cite{Becher:2007ty}
\begin{align}
   \widetilde{S}_{\rm bare}(L,\epsilon) &= 1 + \frac{\alpha_0 C_F}{4\pi}\,e^{-2\epsilon L}
    \left( \frac{4}{\epsilon^2} + \frac{\pi^2}{3} + \frac{4\zeta_3}{3}\,\epsilon 
    + \frac{\pi^4}{40}\,\epsilon^2 \right) \nonumber\\
   &\quad\mbox{}+ \left( \frac{\alpha_0 }{4\pi} \right)^2 e^{-4\epsilon L} 
    \left(  C_F^2 W_F + C_F C_A W_A + C_F T_F n_f W_f \right) ,
\end{align}
with
\begin{align}
   W_F &= \frac{8}{\epsilon^4} + \frac{4\pi^2}{3\epsilon^2}
    + \frac{16\zeta_3}{3\epsilon} + \frac{7\pi^4}{45} \,,\nonumber\\
   W_A &= \frac{11}{3\epsilon^3} + \frac{1}{\epsilon^2}
    \left( \frac{67}{9} - \frac{\pi^2}{3} \right) 
    + \frac{1}{\epsilon} \left(\frac{404}{27}  + \frac{11\pi^2}{18} - 14\zeta_3 \right) 
    + \frac{2428}{81} + \frac{67\pi^2}{54} + \frac{22\zeta_3}{9}- \frac{\pi^4}{3} \,, \nonumber\\
   W_f &= -\frac{4}{3\epsilon^3} - \frac{20}{9\epsilon^2}
    + \frac{1}{\epsilon} \left( - \frac{112}{27} - \frac{2\pi^2}{9} \right) - \frac{656}{81} 
    - \frac{10\pi^2}{27} - \frac{8\zeta_3}{9} \,.
\end{align}
The coft function with two Wilson lines is given by
\begin{align}
   \left\langle \widetilde{\bm{\mathcal{U}}}_{1}(Q\delta\tau,\epsilon) \right\rangle 
   &= 1 + \frac{\alpha_0 C_F}{4\pi}\,e^{-2\epsilon L}
    \left( - \frac{2}{\epsilon^2} - \frac{\pi^2}{2} - \frac{14\zeta_3}{3}\,\epsilon 
    - \frac{7\pi^4}{48}\,\epsilon^2 \right) \nonumber\\
   &\quad\mbox{}+ \left( \frac{\alpha_0}{4\pi} \right)^2 e^{-4\epsilon L} 
    \left( C_F^2 V_F + C_F C_A V_A + C_F T_F n_f V_f \right) ,
\end{align}
where $L=\ln\frac{Q\delta\tau}{\mu}$, and
\begin{align}
   V_F &= \frac{2}{\epsilon^4} + \frac{\pi^2}{\epsilon^2} + \frac{28\zeta_3}{3\epsilon} 
    + \frac{5\pi^4}{12} \,, \nonumber\\
   V_A &= - \frac{11}{6\epsilon^3} - \frac{1}{\epsilon^2}
    \left( \frac{67}{18} + \frac{\pi^2}{6} \right) 
    + \frac{1}{\epsilon} \left(  - \frac{211}{27} - \frac{11\pi^2}{36} + 3\zeta_3 \right)   
    - \frac{836}{81} - \frac{1139\pi^2}{108} - \frac{341\zeta_3}{9}+ \frac{31\pi^4}{90} \,,
    \nonumber\\
   V_f &= \frac{2}{3\epsilon^3} + \frac{10}{9\epsilon^2}
    + \frac{1}{\epsilon} \left( \frac{74}{27} + \frac{\pi^2}{9} \right)  - \frac{374}{81} 
    + \frac{109\pi^2}{27} + \frac{124\zeta_3}{9} \,.
\end{align}
To get this function, we have boosted to the frame where the cone covers the full right hemisphere. In this frame the coft function is the same as the hemisphere soft function $S(\omega_L,\omega_R)$ in the limit $\omega_R\to \infty$, where the energy in the right hemisphere can be arbitrarily large. Taking this limit generates additional singularities, so it needs to be taken before renormalization, using the bare expressions provided in \cite{Schwartz:2014wha}.

We also need the coft-collinear mixing contribution, which involves $L=\ln\frac{Q\delta}{\mu}+\ln\frac{Q\delta\tau}{\mu}$ and reads
\begin{align}
   \left\langle \bm{\mathcal{J}}_{\!\!2}(Q\delta,\epsilon) 
    \otimes \left[\widetilde{\bm{\mathcal{U}}}_{2}(Q\delta\tau,\epsilon) -\bm{1}\right] \right\rangle 
   = \left( \frac{\alpha_0}{4\pi} \right)^2 e^{-2\epsilon L} 
    \left( C_F^2 M_F + C_F C_A M_A\right) ,
\end{align}
with
\begin{align}
   M_F &= - \frac{4}{\epsilon^4} - \frac{6}{\epsilon^3}
    + \frac{1}{\epsilon^2} \left(  - 14 + \frac{2\pi^2}{3} - 12\ln2 \right) 
    + \frac{1}{\epsilon} \left( - 26 - \pi^2 + 10\zeta_3 - 32\ln2   \right) + c_2^{M,F} \,,\nonumber\\
   M_A &= \frac{2\pi^2}{3\epsilon^2}
    + \frac{1}{\epsilon} \left( - 2 + \frac{\pi^2}{2} + 12\zeta_3 + 6\ln^2 2 + 4\ln 2  \right) 
    + c_2^{M,A} \,.
\end{align}  
We have obtained this result from a computation of the relevant diagrams and we computed the constant terms numerically as $c_2^{M,F}=-128.8$ and $c_2^{M,A}=90.53$. Finally, we need the purely collinear contribution, which is obtained as
\begin{equation}
   J^{\rm full}_{\rm bare}(L,\epsilon) = \left\langle \bm{\mathcal{J}}_{\!\!1} \otimes \bm{1}
    + \bm{\mathcal{J}}_{\!\!2} \otimes \bm{1} + \bm{\mathcal{J}}_{\!\!3} \otimes \bm{1} 
    \right\rangle\,,
\end{equation}
with $\langle\bm{\mathcal{J}}_{\!\!1}\otimes\bm{1}\rangle=1$, and has the form  
\begin{align}
   J^{\rm full}_{\rm bare}(L,\epsilon) &= 1 + \frac{\alpha_0 C_F}{4\pi}\,e^{-2\epsilon L}
    \left[ \frac{2}{\epsilon^2} + \frac{3}{\epsilon} + 7 - \frac{5\pi^2}{6} + 6\ln2 
    + \epsilon \left(  14 - \frac{\pi^2}{4} - \frac{44\zeta_3}{3} + 6\ln^2 2 + 14\ln 2 \right)
    \right. \nonumber\\
   &\quad \left. \mbox{}+ \epsilon^2 \left( 28 - \frac{7\pi^2}{12} - \zeta_3 + \frac{41\pi^4}{720}
    - \frac{4\ln^4 2}{3} + 4\ln^3 2 + 14\ln^2 2 + \frac{4\pi^2\ln^2 2}{3} + 28\ln 2
    - \frac{\pi^2\ln 2}{2} \right. \right. \nonumber\\
   &\hspace{1.3cm} \left. \left. \mbox{}- 28\zeta_3 \ln 2 - 32\,\text{Li}_4\Big(\frac{1}{2}\Big) 
    \right) \right] + \left( \frac{\alpha_0}{4\pi} \right)^2 e^{-4\epsilon L} 
    \left( C_F^2 J_F + C_F C_A J_A + C_F T_F n_f J_f \right) ,
\end{align}
with $L=\ln\frac{Q\delta}{\mu}$ and
\begin{align}\label{eq:JfullConst}
   J_F &= \frac{2}{\epsilon^4} + \frac{6}{\epsilon^3}
    + \frac{1}{\epsilon^2} \left( \frac{37}{2} - \frac{5\pi^2}{3} + 12\ln2 \right) 
    + \frac{1}{\epsilon} \left(  \frac{191}{4} - 4\pi^2 - \frac{22\zeta_3}{3}   + 50\ln 2 \right) 
    + c_2^{J,F} \,, \nonumber\\
   J_A &= \frac{11}{6\epsilon^3}
    + \frac{1}{\epsilon^2} \left( \frac{83}{9} - \frac{\pi^2}{2} \right) 
    + \frac{1}{\epsilon} \left( \frac{3985}{108} - \frac{139\pi^2}{36}  - 21\zeta_3 
    - 6\ln^2 2 + 18\ln 2 \right) + c_2^{J,A} \,, \nonumber\\
   J_f &= - \frac{2}{3\epsilon^3} - \frac{28}{9\epsilon^2} 
    + \frac{1}{\epsilon} \left( - \frac{335}{27} + \frac{11\pi^2}{9} - 8\ln 2 \right) 
    + c_2^{J,f} \,.
\end{align}
Note that we did not compute the two-loop coefficients $J_i$ directly but have inferred their divergent parts from the requirement that the cross section is finite. We have obtained numerical values for the finite parts by comparing the result for the cross section to numerical results obtained with the fixed-order event generator {\sc Event2} \cite{Catani:1996vz}. The details of this extraction will be discussed in \cite{longer}.

\subsection*{C: NNLO result for the cone-jet cross section}
\renewcommand{\theequation}{C\arabic{equation}}
\setcounter{equation}{0}

To obtain the NNLO expression for the cone-jet cross section, we now combine the bare ingredients given in the previous section in the form
\begin{equation}
   \widetilde{\sigma}(\tau) = \sigma_0\,H_{\rm bare}\,\widetilde{S}_{\rm bare} 
    \left\langle \widetilde{\bm{\mathcal{U}}}_{1}
    + \bm{\mathcal{J}}_{\!\!2} \otimes \widetilde{\bm{\mathcal{U}}}_{2} 
    + \bm{\mathcal{J}}_{\!\!3} \otimes \bm{1} \right\rangle^2 .
\end{equation}
After coupling renormalization, we obtain a finite result for the Laplace-transformed cross section $\widetilde{\sigma}(\tau)$. The inverse Laplace transformation is then obtained using the simple substitution rules
\begin{equation}
   \ln\tau \to \ln\beta \,, \qquad \ln^2\tau \to \ln^2\beta - \frac{\pi^2}{6} \,.
\end{equation}
It is conventional to choose $\mu=Q$ and write the expansion of the cross section in the form
\begin{align}
   \frac{\sigma(\beta)}{\sigma_0} = 1 + \frac{\alpha_s}{2\pi}\,A(\beta,\delta) 
    + \left(\frac{\alpha_s}{2\pi}\right)^2 B(\beta,\delta) + \dots \,.
\end{align}
The coefficient $A(\beta,\delta)$ was given in the main text. The two-loop coefficient $B(\beta,\delta)$ has the form
\begin{align}\label{Bterms}
   B(\beta,\delta) &= C_F^2 \left[ 
    \left( 32\ln^2\beta + 48\ln\beta + 18 - \frac{16\pi^2}{3}  \right) \ln^2\delta 
    + \left( - 2 + 10\zeta_3  - 12\ln^2 2 + 4\ln 2 \right) \ln\beta \right. \nonumber\\
   &\hspace{1.0cm}\left. \mbox{}+ \left( (8 - 48\ln2) \ln\beta + \frac{9}{2} + 2\pi^2  - 24\zeta_3
     - 36\ln2 \right) \ln\delta + c_2^F \right] \nonumber\\
   &\mbox{}+ C_F C_A \left[ \left( \frac{44\ln\beta}{3} + 11 \right) \ln^2\delta 
    - \frac{2\pi^2}{3} \ln^2\beta 
    + \left(  \frac{8}{3} - \frac{31\pi^2}{18} - 4\zeta_3  - 6\ln^2 2 - 4\ln 2 \right) \ln\beta 
    \right. \nonumber\\
   &\hspace{1.5cm} \left. \mbox{}+ \left( \frac{44\ln^2\beta }{3} 
    + \left( - \frac{268}{9} + \frac{4\pi^2}{3}  \right) \ln\beta - \frac{57}{2} + 12\zeta_3 
    - 22\ln 2 \right) \ln\delta + c_2^A \right] \nonumber\\
   &\mbox{}+ C_F T_F n_f \left[ \left( - \frac{16\ln\beta}{3} - 4 \right) \ln^2\delta 
    + \left( - \frac{16}{3} \ln^2\beta + \frac{80\ln\beta}{9} + 10 + 8\ln 2 \right) \ln\delta 
    + \left( - \frac{4}{3}  + \frac{4\pi^2}{9} \right) \ln\beta + c_2^f \right] .
\end{align}
The quantities $c_2^F$, $c_2^A$ and $c_2^f$ are directly related to the unknown constants $c_2^{J,F}$, $c_2^{J,F}$ and $c_2^{J,f}$ in \eqref{eq:JfullConst}. We have determined them numerically by running the {\sc Event2} generator at low values of $\delta$ and $\beta$. Subtracting the known logarithmic structure exhibited in \eqref{Bterms}, we can then fit for the numerical values of the constants and obtain
\begin{equation}
   c_2^F = 17.1^{+3.0}_{-4.7} \,, \qquad
   c_2^A = -28.7^{+0.7}_{-1.0} \,, \qquad
   c_2^f = 17.3^{+0.3}_{-9.0} \,.
\end{equation}
The uncertainty on the last constant is fairly large due to numerical instabilities \cite{longer}.

\end{document}